\documentclass{aastex62}

\accepted{by ApJ, September 6, 2018}

\shorttitle{Solar Eruption Associated with X9.3 flare}
\shortauthors{Inoue et al.}

\begin{document}

\title{Magnetohydrodynamic Modeling of a Solar Eruption Associated with X9.3 Flare Observed in Active Region 12673}

\correspondingauthor{Satoshi Inoue}
\email{inosato@nagoya-u.jp}

\author[0000-0002-0786-7307]{Satoshi Inoue}
\affil{Institute for Space-Earth Environmental Research (ISEE), Nagoya University Furo-cho, Chikusa-ku, Nagoya, 464-8601, Japan}

\author{Daikou Shiota}
\affiliation{National Institute of  Information and Communications Technology (NICT), 4-2-1, Nukui-Kitamachi, Koganei, Tokyo 184-8795, 
                Japan}
\affiliation{Institute for Space-Earth Environmental Research (ISEE), Nagoya University Furo-cho, Chikusa-ku, Nagoya, 464-8601, Japan}

\author{Yumi Bamba}
\affiliation{Institute of Space and Astronautical Science (ISAS)/Japan Aerospace Exploration Agency (JAXA) 
                3-1-1 Yoshinodai, Chuo-ku, Sagamihara, Kanagawa 252-5210}

\author{Sung-Hong Park}
\affiliation{Institute for Space-Earth Environmental Research (ISEE), Nagoya University Furo-cho, Chikusa-ku, Nagoya, 464-8601, Japan}



\begin{abstract}
 On SOL2017-09-06 solar active region 12673 produced an X9.3 flare which is regarded as largest to occur in solar cycle 24. In this work
 we have preformed a magnetohydrodynamic (MHD) simulation in order to reveal the three-dimensional (3D) dynamics of the 
 magnetic fields associated with the X9.3 solar flare.  We first performed an extrapolation of the 3D magnetic field based on the observed 
 photospheric magnetic field prior to the flare and then used it as the initial condition for an MHD simulation. Consequently, 
 the simulation showed a dramatic eruption. In particular, we found that a large coherent flux rope composed of highly twisted magnetic field 
 lines is formed during the eruption. A series of small flux ropes are found to lie along a magnetic polarity inversion line prior to the flare. 
 Reconnection occurring between each small flux rope during the early stages of the eruption forms the large and highly twisted flux rope.
 Furthermore, we found a writhing motion of the erupting flux rope. The understanding of these dynamics is important in increasing the
 accuracy of space weather forecasting. We report on the detailed dynamics of the 3D eruptive flux rope and discuss the possible
 mechanisms of the writhing motion.
  \end{abstract}

\keywords{Sun:Magnetic Fields, Sun:Solar Flares, Sun:Coronal Mass Ejections}


\section{Introduction} \label{sec:intro}
 Highly energetic phenomena observed in the solar corona are mainly driven by coronal magnetic fields(\citealt{2002A&ARv..10..313P}). 
 These magnetic fields are distorted by the plasma flows associated with photospheric motions {\it e.g.}, the shear motion and 
 converging motion(\citealt{1989ApJ...343..971V}, \citealt{2003ApJ...585.1073A}, \citealt{2013ApJ...778...13P}, 
 \citealt{2018ApJ...860..163W}, \citealt{2018SoPh..293....114}). The magnetic fields are largely deformed from their potential state,
 resulting in the accumulation of free magnetic energy which is eventually  released({\it e.g.}, \citealt{2014IAUS..300..184A}). This energy 
 release is often observed in the form of solar flares, resulting in the heating of plasma and acceleration of 
 particles(\citealt{2015SoPh..290.3425J}, \citealt{2017LRSP...14....2B}). Eruptive magnetic fields and accompanying plasma sometimes grow
 into coronal mass ejections(CMEs:  \citealt{2015SoPh..290.3457S}, \citealt{2017LRSP...14....5K}) which sometimes propagate through 
 interplanetary space and interact with the Earth's magnetosphere, resulting on occasions in large disturbances to the electromagnetic
  environment of the near-Earth space.
 
 The magnetic flux rope(MFR), which is a bundle of helically twisted magnetic field lines which carry the current density, is 
 a key structure when considering these solar eruptions. This is because the MFR is believed to be one of the possible pre-eruptive 
 configurations which is destabilized at the initiation of the eruption and becomes part of the core of a CME(\citealt{2000JGR...10523153F}, 
 \citealt{2017PhPl...24i0501C}). However, the origin and dynamics of MFRs are not fully understood. For instance, there is large gap between 
 the estimation of the magnetic twist observed in the lower corona and interplanetary space({\it e.g.}, \citealt{2017NatCo...8.1330W}). 
 Several simulations, {\it e.g.}, \cite{2006ApJ...637L..65G} and \cite{2017ApJ...850...95S} suggest that the twist accumulation would be 
 possible during an eruption through tether-cutting reconnection(\citealt{2001ApJ...552..833M}). 
 
 Furthermore, eruptive MFRs sometimes show rotation and deflection(\citealt{2017SoPh..292...78K}). These are key properties to 
 determine the direction of the magnetic field observed in vicinity of the Earth, which is important to space weather forecasts.
 Although the writhing motion contributing to the rotation of MFR is sometimes observed in the lower corona ({\it e.g.,} 
 \citealt{2003SoPh..214..313R}, \citealt{2005ApJ...628L.163W}, \citealt{2010A&A...516A..49T}, \citealt{2012SoPh..281..137K},  and
 \citealt{2014ApJ...782...67Y}), quantitative analysis is difficult because no information about 3D coronal magnetic field is available, which 
 prevents us from fully understanding of the dynamics of MFRs. Recently, data-constrained and -driven magnetohydrodynamic (MHD) 
 simulations have been vigorously performed({\it e.g.}, \citealt{2015ApJ...803...73I}, \citealt{2016NatCo...711522J}, 
 \citealt{2017ApJ...842...86M}, \citealt{2018Natur.554..211A} and \citealt{2018ApJ...860...96P}). Because these simulations clarify the 3D 
 dynamics of the magnetic field bounded by the observed photospheric magnetic field, these have the potential to bring about more 
 quantitative interpretations. 
 
 In this study, in order to clarify the dynamics of eruptive MFRs, we perform a data-constrained MHD simulation of a solar eruption. In 
 September 2017, solar active region (AR) 12673 ({\it e.g.,}\citealt{2017ApJ...849L..21Y},  \citealt{2018ApJ...856...79Y}) rapidly grew over the 
 course of a few days. The AR became very active from September 4th, and produced many M\,- and C\,-\,class flares during two days, 
 eventually resulting in the production of an X2.2 flare at 08:57 UT on September 6th. Furthermore, approximately three hours later, at 11:53 
 UT, an X9.3 flare was observed which is recorded as the largest flare of solar cycle 24. Since this flare is associated with a geo-effective 
 CME, it is a good case to investigate the dynamics of the geo-effective solar MFR.  The data available for this AR is rich, nevertheless, it is
 hard to determine the 3D magnetic structure as current observations do not allow direct measurements of the coronal magnetic fields. To 
 overcome this issue, nonlinear force-free field extrapolations (NLFFF:\citealt{2012LRSP....9....5W}, \citealt{2016PEPS....3...19I}, 
 \citealt{2017ScChE..60.1408G}), where 3D magnetic fields are extrapolated from the observed photospheric magnetic field under a 
 force-free approximation, are employed.  Eventually, in order to investigate their dynamics during the X9.3 flare, we carry out MHD 
 simulations using the NLFFF as the initial condition.  
 
 The rest of this paper is constructed as follows: the observations and numerical method are 
 described in section 2; results are presented in section 3; and finally, important discussions arising from our findings are summarized in 
 section 4.  \\

    \section{Observations $\&$ Method} 
    \subsection{Observations}
    As shown in Fig.\ref{f1}(a), AR12673 was very active, with M-class flares being observed 4 times from September 5th. Eventually the 
    X2.2 and X9.3 flares were observed to occur on September 6th.  Figure 1(b) shows the extreme ultraviolet (EUV) image of whole sun 
    observed by the Atmospheric Imaging Assembly (AIA) (\citealt{2012SoPh..275...17L}) on board {\it Solar Dynamics Observatory (SDO}: 
    \citealt{2012SoPh..275....3P}) in 131 \AA.  AR 12673 is highlighted by the marked box in the lower western quadrant of the sun.  
    Figure \ref{f1}(c) shows the photospheric vector magnetic field of AR12673, observed at 08:36 UT on September 6th taken from 
    Helioseismic and Magnetic Imager (HMI: \citealt{2012SoPh..275..207S}) onboard {\it SDO}. This observation is shown in Space 
    weather HMI Active Region Patch(SHARP) format (\citealt{2014SoPh..289.3549B}). This was taken approximately 20 minuets before the 
    X2.2 flare, from which the presence of sheared magnetic field lines along the polarity inversion line (PIL) can be identified. In this study, 
    this photospheric magnetic field was used to perform a NLFFF extrapolation.

    \subsection{Numerical Methods}
    \subsubsection{Nonlinear Force-Free Field Extrapolation}
    We first perform the NLFFF extrapolation using the MHD relaxation method described in \cite{2014ApJ...780..101I} and 
    \cite{2016PEPS....3...19I}. The photospheric vector magnetic field, as shown in Fig. \ref{f1}c, is set as the bottom boundary where it is 
    preprocessed according to \cite{2006SoPh..233..215W}.  To perform the NLFFF extrapolation and the MHD simulation we solve the 
    following equations:
   \begin{equation}
   \rho = |{\bf B}|,
   \label{den_eq}
   \end{equation}

    \begin{equation}
    \frac{\partial {\bf v}}{\partial t} 
                         = - ({\bf v}\cdot{\bf \nabla}){\bf v}
                           + \frac{1}{\rho} {\bf J}\times{\bf B}
                           + \nu_i{\bf \nabla}^{2}{\bf v},
   \label{eq_of_mo}    
   \end{equation}

  \begin{equation}
  \frac{\partial {\bf B}}{\partial t} 
                        =  {\bf \nabla}\times({\bf v}\times{\bf B}
                        -  \eta_{\rm i}{\bf J})
                        -  {\bf \nabla}\phi, 
  \label{in_eq}
  \end{equation}

  \begin{equation}
   {\bf J} = {\bf \nabla}\times{\bf B},
  \label{Am_low}
  \end{equation}
  
  \begin{equation}
  \frac{\partial \phi}{\partial t} + c^2_{\rm h}{\bf \nabla}\cdot{\bf B} 
    = -\frac{c^2_{\rm h}}{c^2_{\rm p}}\phi,
  \label{div_eq}
  \end{equation}
  where the subscript i of $\nu$ and $\eta$ corresponds to different values used in NLFFF or MHD, respectively. $\rho$ is pseudo plasma 
  density, ${\bf B}$ is the magnetic flux density, ${\bf v}$ is the velocity, ${\bf J}$ is the electric current density, and $\phi$ is the convenient 
  potential to reduce errors derived from ${\bf \nabla}\cdot {\bf B}$ (\citealt{2002JCoPh.175..645D}), respectively. The pseudo density is 
  assumed to be proportional to $|{\bf B}|$ in order to ease the relaxation by equalizing the {\bf Alfv\'en} speed in space.  In these equations, 
  the length, magnetic field, density, velocity, time, and electric current density are normalized by   
  $L^{*}$ = 244.8   Mm,  
   $B^{*}$ = 2500 G, 
   $\rho^{*}$ = $|B^{*}|$,
   $V_{\rm A}^{*}\equiv B^{*}/(\mu_{0}\rho^{*})^{1/2}$,    
   where $\mu_0$ is the magnetic permeability,
   $\tau_{\rm A}^{*}\equiv L^{*}/V_{\rm A}^{*}$, and     
   $J^{*}=B^{*}/\mu_{0} L^{*}$,  
   respectively.
   $\nu_{\rm NLFFF}$ is a viscosity fixed by $1.0 \times 10^{-3}$, and the coefficients $c_h^2$, $c_p^2$ in Equation (\ref{div_eq}) also fixed
   the constant value, 0.04 and 0.1, respectively.  The resistivity in the NLFFF calculation is given as 
   $\eta_{\rm NLFFF} = \eta_0 + \eta_1 |{\bf J}\times{\bf B}||{\bf v}|/{\bf |B|}$ where $\eta_0 = 5.0\times 10^{-5}$ and 
   $\eta_1=1.0\times 10^{-3}$ in non-dimensional units. The second term is introduced to accelerate the relaxation to the force-free 
   state, particularly in regions of weak field. 
    
  A potential field is employed as the initial state in the NLFFF calculation, which is extrapolated from the observed $B_z$ using the Green 
  function method(\citealt{1982SoPh...76..301S}). For both calculations, the density is initially given by $\rho = |{\bf B}|$ and the velocity is 
  set to zero everywhere inside the numerical domain. At the boundaries except the bottom, the magnetic fields are fixed to be potential fields.
  The bottom boundary is fixed according to  ${\bf B}_{bc} = \zeta {\bf B}_{obs} + (1-\zeta) {\bf B}_{pot}$ where ${\bf B}_{bc}$  is the 
  horizontal  component which is determined by a  linear combination of the observed magnetic field (${\bf B}_{obs}$) and the potential 
  magnetic field (${\bf B}_{pot}$). $\zeta$ is a coefficient ranging from 0 to 1. When $R=\int |{\bf J}\times{\bf B}|^2$dV, which is calculated 
  over the interior of the computational domain, falls below a critical value denoted by $R_{min}$ during the iteration, the value of the 
  parameter $\zeta$ is increased to $\zeta = \zeta + d\zeta$. In this paper, $R_{min}$ and d$\zeta$ have the values $1.0 \times 10^{-2}$ 
  and 0.02, respectively. If $\zeta$ becomes equal to 1, ${\bf B}_{bc}$ is completely consistent with the observed data. This process 
  can suppress large discontinuities produced between the bottom and inner domain.  The velocity is fixed to zero at all boundaries and 
  the von Neumann condition $\partial /\partial n$=0 is imposed on $\phi$ in both calculations. We further controlled the velocity as 
  follows. If the value of $v^{*}(=|{\bf v}|/{|\bf v}_{\rm A}|)$ becomes larger than $v_{max}$(here set to 0.02), then the  velocity is modified as 
  follows: ${\bf v} \Rightarrow (v_{max}/ v^{*}) {\bf v}$. These process can also suppress large discontinuities produced between the bottom 
  and inner domain.
  
   \subsubsection{Magnetohydrodynamic Simulation}  
    Next we perform the MHD simulation using the NLFFF as an initial condition. The solved equations are identical to those used in 
    the NLFFF extrapolation. The pseudo density  is used again in the MHD simulation. The reason for this is that the dynamics in the lower 
    corona are not so dissimilar from those produced using a mass equation, as shown by \cite{1999ApJ...518L..57A} and 
    \cite{2014ApJ...788..182I}. Regarding differences resulting from the NLFFF calculation, we set a uniform resistivity by fixing 
    $\eta_{\rm MHD}=1.0 \times 10^{-5}$ and $\nu_{\rm mhd} =1.0 \times 10^{-4}$. At the boundaries, only the normal 
    component of the magnetic field($B_{\rm n}$) is fixed, while the horizontal components vary according to the dynamics {\it i.e.}, they are 
    determined by the induction equation. The velocity limiter used for the NLFFF extrapolation is not used for the MHD simulation.
    
   For both calculations, the numerical domain has dimensions of $244.8 \times 158.39 \times 195.84 (\rm Mm^3)$, or $1.0 \times 0.647 
   \times 0.8$ in non-dimensional units. The region is divided into $340 \times 220 \times 272$ grid points resulting in a $2 \times 2$ binning 
   of the original photopsheric vector magnetic field.   

    \section{Results} \label{sec:floats}

  \begin{figure}
  \epsscale{.9}
  \plotone{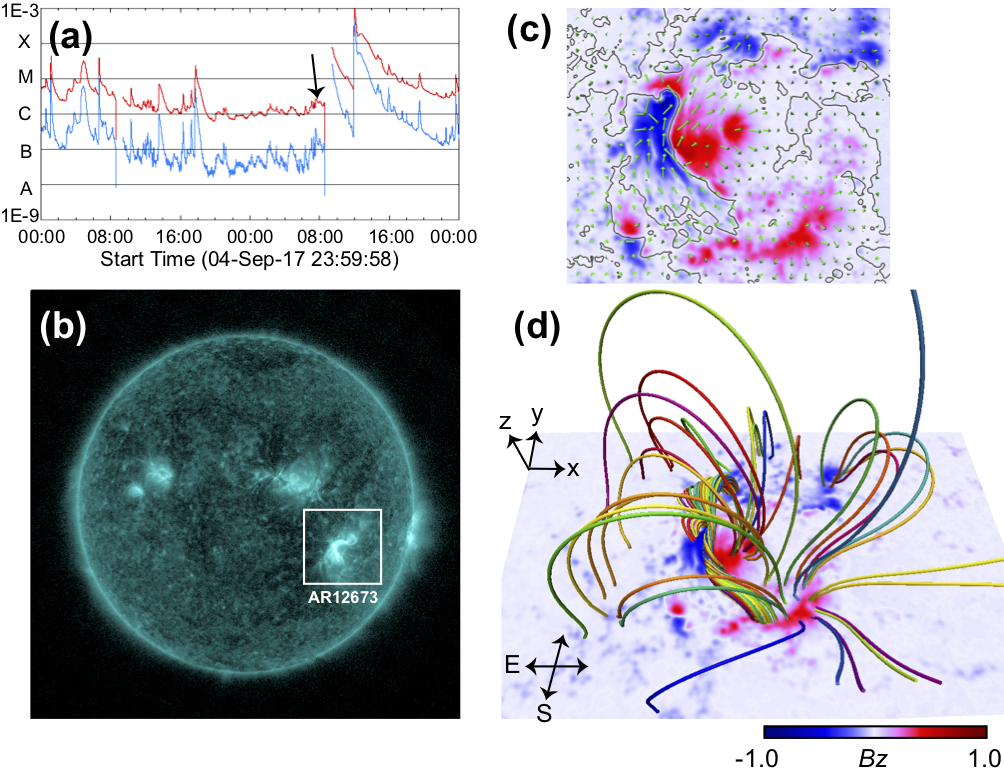}
  \caption{
          (a) Time profile of the X-ray flux measured by the GOES 12 satellite between 5th-6th September 2017. 
               The solar X-ray emission in the 1$-$8 \AA \ and  $0.5-4.0$ \AA \ passband are plotted in red and blue respectively. Several 
               M-class and X-class  flares, one of which is {\bf X9.3 solar flare}, were produced in this AR.  
          (b) EUV image of whole sun observed in {\it AIA} 131 \AA\, in which AR12673 is surrounded by a white square. 
          (c) The photospheric magnetic field taken at 08:36 UT on 6th September, approximately 20 minutes before the X2.2 flare.
               The red/blue image shows the $B_z$ component of the magnetic field, while the arrows show to the horizontal component.
          (d) Three-dimensional magnetic fields extrapolated from the photospheric magnetic field shown in (c), under the NLFFF assumption.
               The magnetic field lines are plotted on $B_z$ distribution set at bottom.       
               }
  \label{f1}
  \end{figure}

\begin{figure}
  \epsscale{1.}
  \plotone{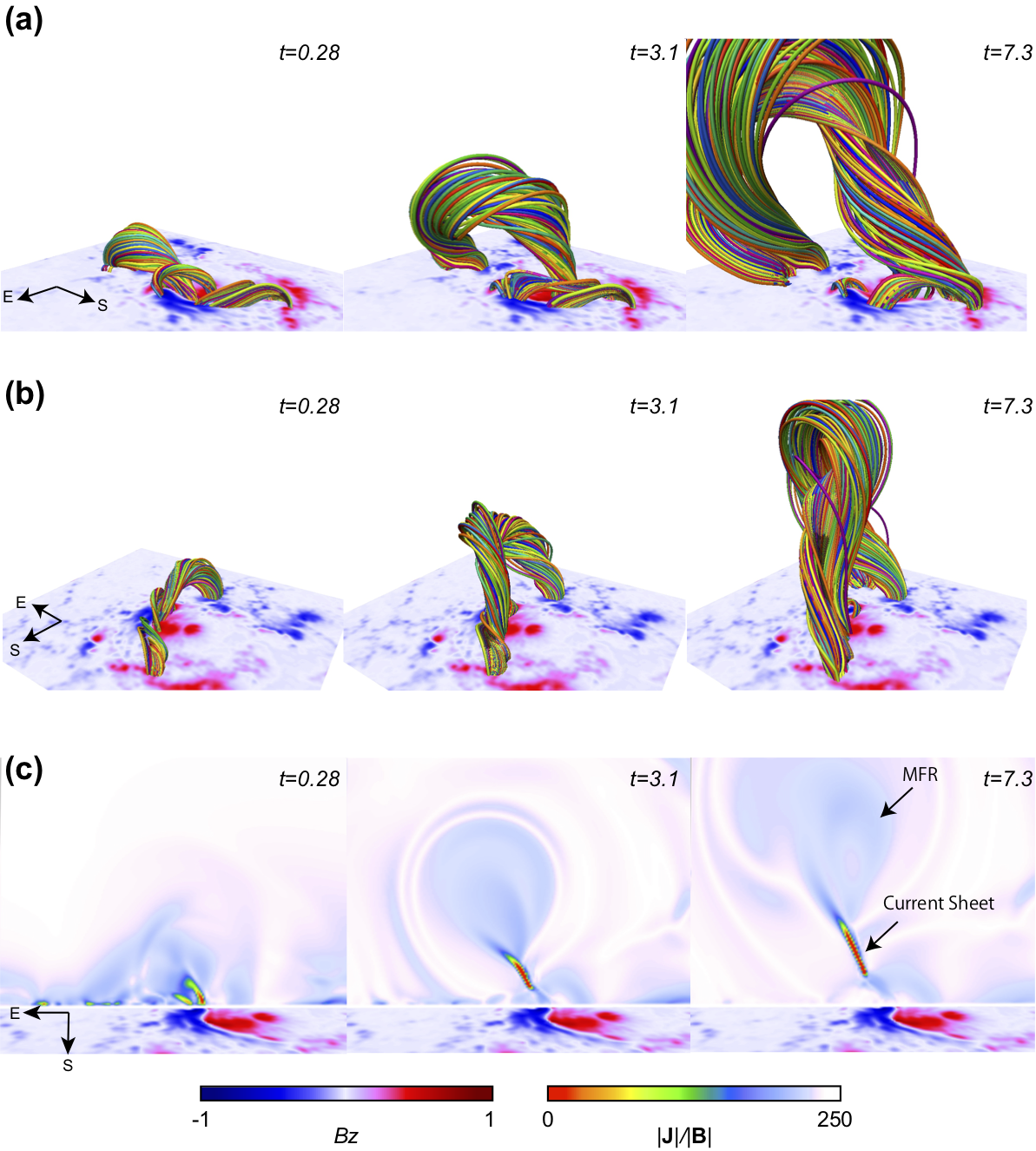}
  \caption{
               Temporal evolution of the formation and dynamics of the eruptive MFR.  Upper(a) and middle(b) panels show the field lines from 
               different angle. E and S mean directions of "East" and "South". (c) Temporal evolution of $|{\bf J}|/|{\bf B}|$ distribution plotted on
                $x-z$ plane at $y=0.38$ in a viewpoint from the south. 
          }
  \label{f2}
  \end{figure}
  
\begin{figure}
  \epsscale{.75}
  \plotone{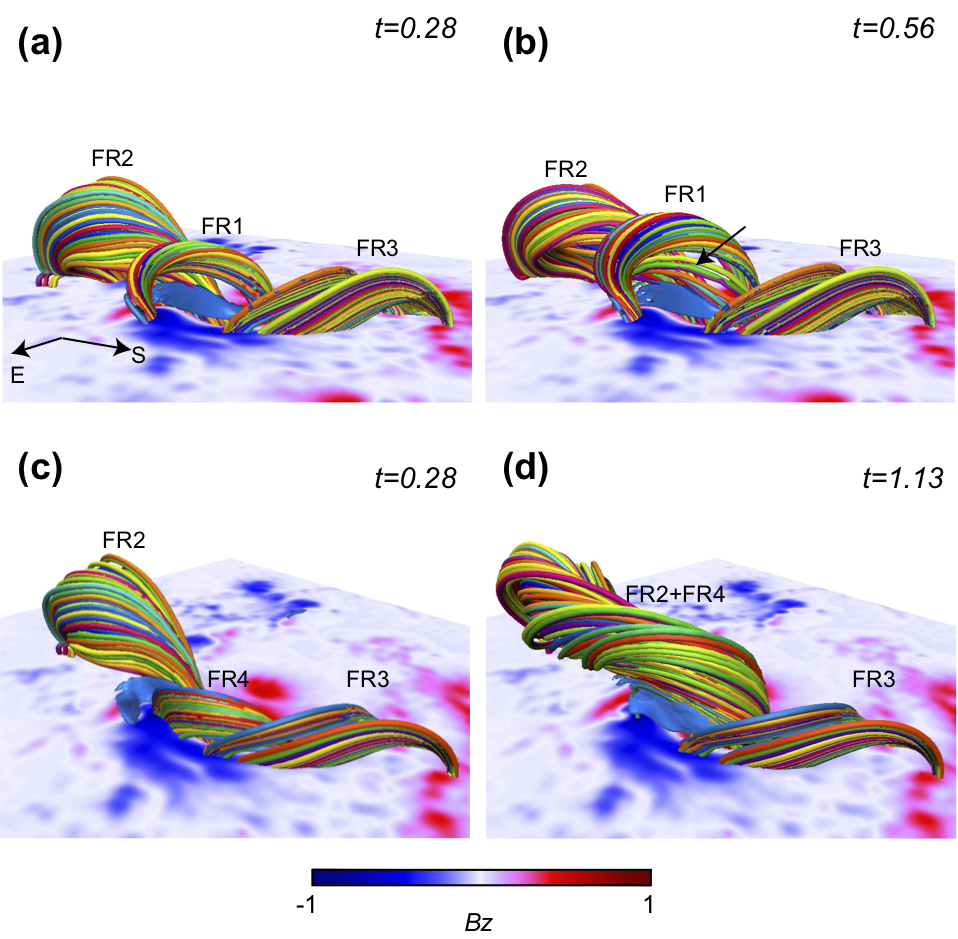}
  \caption{
               (a)-(b) Small flux ropes named as FR1, FR2 and FR3 at $t=$0.28 and $t=$0.56, respectively. Blue surface corresponds to 
               isosurface of the current density $|{\bf J}|=50$.
               (c)-(d) FR4 is plotted, which is formed under FR1, in addition to FR2 and FR3 at $t=$0.28 and 1.13, respectively. 
                Note that FR1 is not plotted. Blue surface corresponds to isosurface of the current density $|{\bf J}|=$35.
               }
  \label{f3}
  \end{figure} 
  
\begin{figure}
  \epsscale{1.}
  \plotone{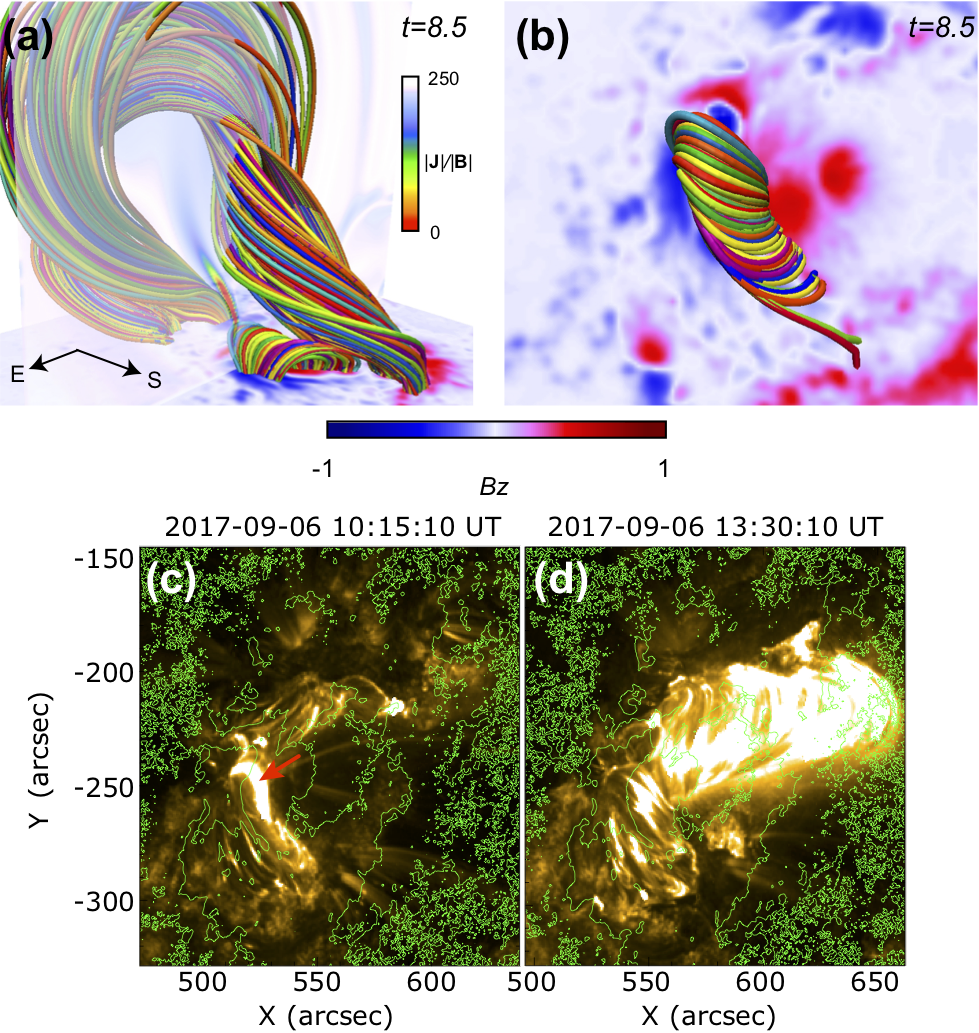}
  \caption{
               (a) A snapshot of the 3D field lines at $t$=8.5 with $|{\bf J}|/|{\bf B}|$ plotted on $x-z$ plane at $y=0.38$ and $B_z$ distribution at 
                    the bottom. 
               (b) Enlarged view of the post-flare loops from top. 
               (c)-(d) AIA 171 \AA \ images taken at 10:15 UT and 13:30 UT, respectively, before and after the X9.3 flare. The green line 
                   corresponds to the PIL and red arrow points out the representative non-potential field lying the PIL. Note that these  images are 
                   taken after the X2.2 flare. 
               }
  \label{f4}
  \end{figure}

  \begin{figure}
  \epsscale{1.}
  \plotone{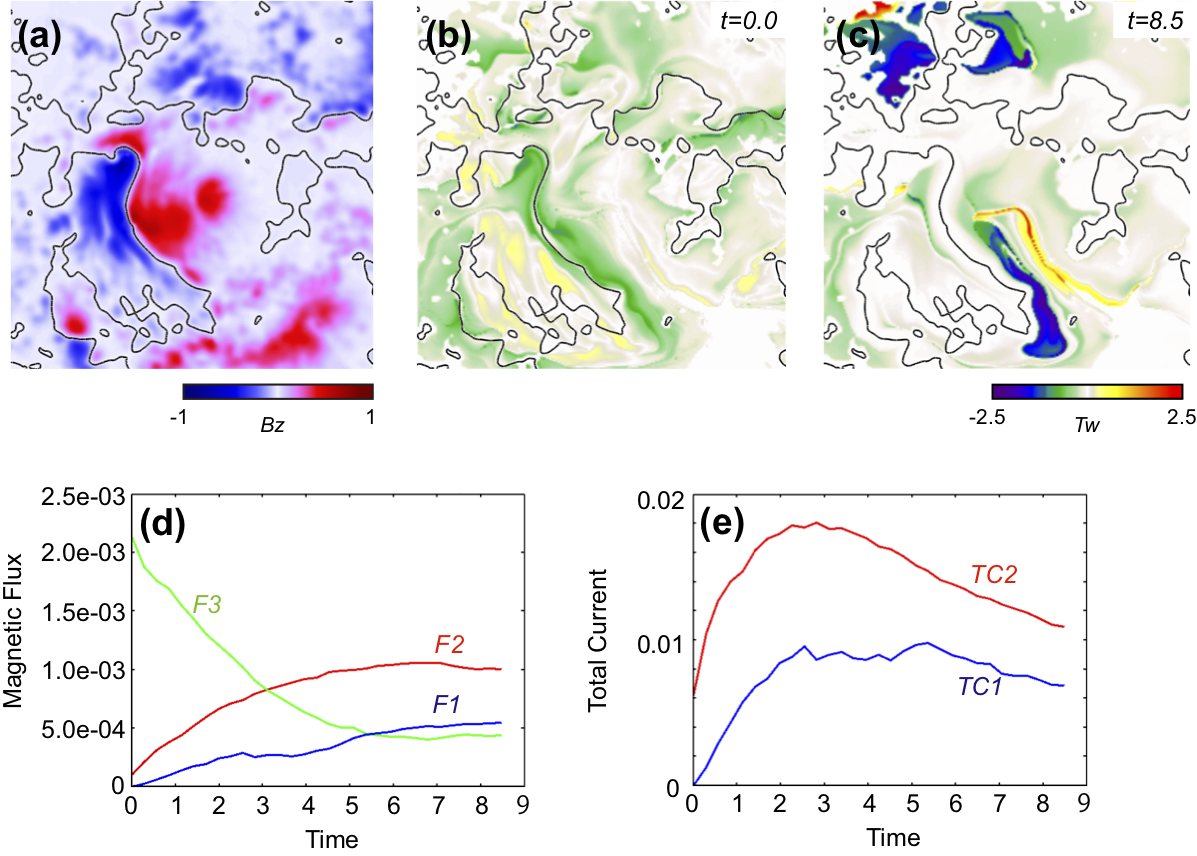}
  \caption{
                (a)       The distribution of $Bz$ with PIL in black.
                (b)-(c) The $T_w$, obtained from the simulation, mapped on the photosphere at $t$=0, and $t$= 8.5, respectively.  
                           The region is identical to one in (a) and the black line corresponds to the PIL.  
                (d) Temporal evolution of the magnetic flux, F1, F2, and F3 respectively, measured in each $T_w$ range of  -1.5$\geq T_w$, 
                     -1.0$\geq T_w$, and -0.5 $\geq T_w \geq$-1.0, in blue, red  and green, respectively.                
                (e) Temporal evolution of the total currents (TC1 and TC2) of the twisted field lines satisfying $T_w \leq -1.5$ and $T_w \leq -1.0$, 
                      respectively. Blue and red lines correspond to TC1 and TC2, respectively. 
                }
  \label{f5}
  \end{figure}
    
  \begin{figure}
  \epsscale{1.}
  \plotone{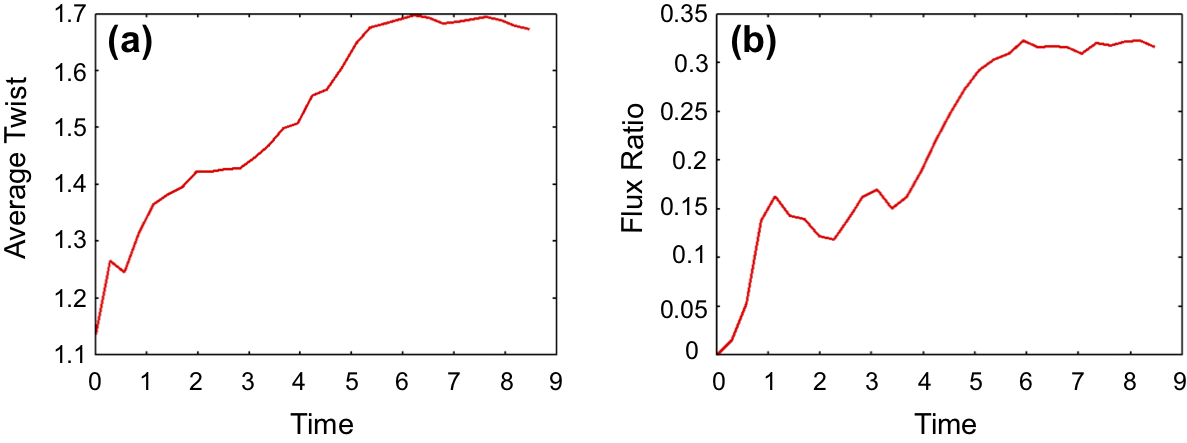}
  \caption{ 
              (a) Temporal evolution of the average twist $<T_w>$ defined in Eq.(\ref{ave_tw}).    
              (b) Temporal evolution of the ratio of the magnetic flux defined in Eq. (\ref{flux_ratio}).
              }
  \label{f6}
  \end{figure}

  \begin{figure}
  \epsscale{.6}
  \plotone{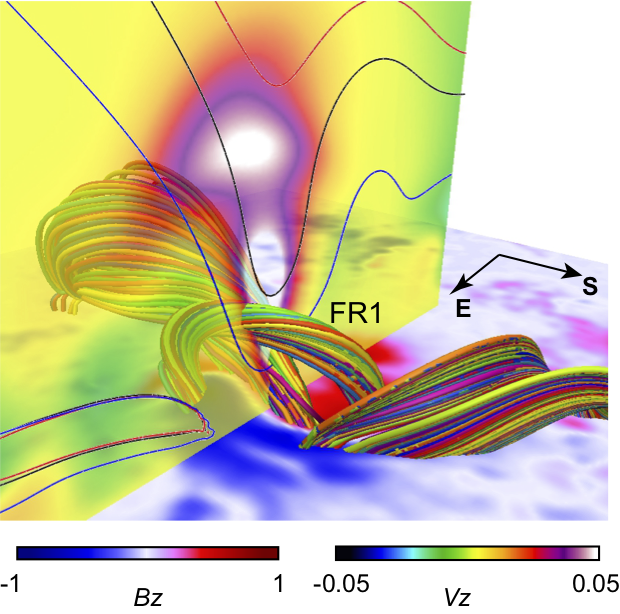}
  \caption{ A snapshot of 3D magnetic fields at $t$=0.28 at which the eruption starts. Vertical cross section shows the $V_z$ distribution 
                at Y=0.34 while the blue, black and red lines correspond to a contour of a decay index value $n$= 1.0, 1.5, and 1.7, 
                 respectively.              
              }
  \label{f7}
  \end{figure}

     Figure 1(d) shows the NLFFF extrapolated from the photospheric magnetic field prior to the X2.2 flare shown in Fig.\ 1(c). 
    The extrapolation reveals twisted magnetic field lines along the north-south direction, as expected from the photospheric magnetic field.
    Next we show results of the MHD simulation using the NLFFF as the initial condition. We first show the overview of the dynamics of the 
    magnetic field(refer also to supplemental online). The left panel of Fig. \ref{f2}(a) shows some of the small MFRs located along the PIL 
    prior to the eruption. After the initiation of the eruption, a large coherent MFR composed of highly twisted lines is formed through 
    reconnection between the small MFRs(middle and right panels). This process is similar to those found in \cite{2018NatCo...9..174I}. 
    In Fig.\ref{f2}(b) which shows the Fig.\ref{f2}(a) 3D field lines from a different angle, the eruptive MFR exhibits a writhing motion. 
    Fig.\ref{f2}(c) shows the temporal evolution of $|{\bf J}|/|\bf B|$ plotted on the $x-z$ plane set at $y=0.38$. The $|{\bf J}|/|\bf B|$ value is 
    enhanced inside the MFR and a current sheet is formed below it, similar to the CSHKP standard model of 
    flares(\citealt{1964NASSP..50..451C}, \citealt{1966Natur.211..695S}, \citealt{1974SoPh...34..323H}, \citealt{1976SoPh...50...85K}). 
    We further found the MFR launches to an eastward direction. 

     A more detailed description of the MFRs and their dynamics at early stages of the simulation is shown in Fig.\ref{f3}. Along the PIL at 
    $t=$0.28, we found three small MFRs named as FR1, FR2, and FR3. They are the same field lines shown in the left panel of 
    Fig. \ref{f2}(a). FR1 expands upward at $t=$0.56. The initiation mechanism of the FR1 will be discussed later. The black arrow in 
    Fig. \ref{f3}b($t=$0.56) shows several elongated field lines of FR2, that now extend also below FR1. These field lines are located 
    above the strong current that develops underneath FR1 as the latter expands. This suggests that reconnection occurs in the strong current
    region. Figs.\ref{f3}(c)-(d) show additional field lines, named as FR4, which are located under FR1(not plotted here). The field lines of FR4 
    and FR2 reconnect through the strong current region and create a bigger MFR({\it i.e.,}FR2+FR4 in Fig.\ref{f3}(d)). This MFR eventually 
    reconnects with the FR3 forming a large MFR, shown in Fig.\ref{f4}(a).
    
     Figure \ref{f4}(a) shows selected magnetic field lines at $t=8.5$ with the ${\bf |J|}/{\bf |B|}$ plotted on the $x-z$ plane. 
    We can see the post flare loops formed below the vertical current sheet and also the eruptive MFR. Fig. \ref{f4}(b) is a top 
    and enlarged view of the post-flare loops, showing that the region close to the PIL is mostly dominated by the post-flare loops.  Although 
    the X9.3 flare, which took place approximately 3 hours after the X2.2 flare, requires highly twisted field lines to be energized, most of the 
    low-lying twisted lines in our simulation have been convert to the post-flare loops. However, from observations of AIA171 \AA \ shown in
    Figs.\ref{f4}(c) and (d), we can see that the sheared magnetic field highlighted by red arrow remains even after the X2.2 flare whilst the 
    post-flare loops appear there after the X9.3 flare. Therefore we suggest that our simulation shows both the X2.2 and X9.3 flares. The X2.2
    flare could be associated with the rise of a relatively small-scale MFR(FR1) early in the simulation($t \approx 0.5$), while the X9.3 flare 
    could be associated with eruption of the large-scale MFR(FR2+FR3+FR4) found later in the simulation. Note that the 3D NLFFF 
    reconstructed from the photospheric vector magnetic field data approximately 20 minutes before the X2.2 flare was used for the initial state
    of the simulation.
       
    For more detailed understanding, we estimate the magnetic twist according to \cite{2006JPhA...39.8321B}
    \begin{equation}
    T_{\rm w}=\frac{1}{4\pi} \int \frac{{\bf \nabla}\times{\bf B}\cdot{\bf B}}{|{\bf B}|^2}dl,
    \end{equation}
    where {\it dl} is a line element. We computed $T_w$ for the field lines rooted in the region shown in Fig.\ref{f5}(a). Figure \ref{f5}(a) shows 
    the $B_z$ distribution obtained at same time shown in Fig.\ref{f1}(c). The value of twist at $t$=0.0 and $t=$8.5,  mapped on the 
    photosphere, are plotted in Figs.\ref{f5}(b) and (c), respectively. Although the highly twisted lines are formed at $t$=8.5 and the writhing 
    motion is found during the eruption, the initial state is stable to kink instability (KI) because the twist value does not reach the required 
    threshold ($T_w \ge$ 1.75) (\citealt{2004A&A...413L..27T}). At $t=$0.85, we found that the value of $T_w$ decreases in the region close 
    to the PIL, where the post-reconnection arcade of Fig. \ref{f4}(b) is rooted.
    Figure \ref{f5}(d) shows the temporal evolutions of the magnetic flux {\it F1}, {\it F2}, and {\it F3} of regions that have $T_w$ within the 
    range of $-1.5 \geq T_w$, $-1.0 \geq T_w$ and $-0.5 \geq T_w \geq -1.0$, respectively. We found that {\it F1} and {\it F2} constantly
    increase as time passes, whilst {\it F3} constantly decreases. {\it F1}, {\it F2}, and {\it F3} all saturate after $t \approx$5.3. This result 
    suggests that the highly twisted MFR is built up from the reconnection of moderately twisted lines(in the twist range of 
    $-0.5 \geq T_w \geq -1.0$, and after $t \approx$5.3,  it expands upwards without further increase of its twist. We further plot  the temporal
    evolutions of the total currents (TC1 and TC2) of the field lines satisfying $T_w \leq  -1.5 $ and $T_w \leq -1.0$, respectively. 
    We define the `total current' as the total current vertically crossing the section ({\bf S}) of the MFR, {\it i.e.}, it is described as 
    $|\int {\bf J} \cdot d{\bf S}|$ where d{\bf S} is the cross-sectional element of the MFR. In Fig.\ref{f5}(e) both TC1 and TC2 increase to 
    $t \approx$3. Afterwards, TC2 decreases, while TC1 remains constant up to $t \approx$5, and then decreases. In this period, the highly 
    twisted MFR is created.  However, both TC1 and TC2 are found to decrease during the late stages of the simulation, in which the MFR 
    travels upward whilst expanding.     
    \section{Discussion}
    In this section we discuss possible mechanisms for the writhing motion of the erupting MFR and also for the initiation of the flare. Several 
    possibilities can be raised to explain the writhing motion of the MFR (\citealt{2012SoPh..281..137K}) one of which is KI. The threshold of KI 
    is derived from a winding number N of a field line around magnetic axis of the MFR(\citealt{1979SoPh...64..303H}). The criteria 
    corresponds to N=1.25 for a cylindrical MFR(\citealt{1979SoPh...64..303H}) and N=1.75 for  a semi-torus type MFR 
    (\citealt{2004A&A...413L..27T}). Note that these values are obtained from a linear stability analysis which are not exactly  applied to the 
    MFR in dynamic state. We therefore use them as an indicator. We define the average twist as follows, 
    \begin{equation}
     <T_w>= \frac{1}{(N_xN_y)}\sum_{i,j, (T_w \leq -1.0)}^{N_x, N_y} |T_w(x_i,y_j)|,  
        \label{ave_tw}
    \end{equation}
    where  $N_x$ and $N_y$ are grid number in the calculated area in the $x$ and $y$ directions, respectively. Note that the averaged 
    twist is calculated in the region shown in Fig. \ref{f5}(a). We select field lines with $T_w \leq -1.0$ as being the ones consisting the new 
    MFR created during the eruption.  Figure \ref{f6}(a) shows that the maximum $<T_w>$  does not reach the value 1.7. We further plot the 
    temporal evolution for the ratio of the magnetic flux described as follows, 
    \begin{equation}
      F_R= \frac{\int_{T_w \leq -1.75} |B_z| dS}{\int_{T_w \leq -1.0} |B_z| dS}, 
      \label{flux_ratio}
    \end{equation}
    where $dS$ is a surface element on the photosphere. Figure \ref{f6}(b) shows that the ratio of the photospheric magnetic flux of  the field
    lines  with $T_w \leq -1.75$ over the photospheric flux of field lines with $T_w \leq -1$ is 30 $\%$. We therefore suggest that KI is not the
    main driver of the writhing motion of the MFR. Another possibility is the ${\bf J}\times{\bf B}$ force acting on the eruptive
    MFR. According to \cite{2007ApJ...670.1453I}, even if the field lines are initially aligned with the MFR current( {\it i.e.}, no 
    force acting on the MFR), once the MFR starts to move upwards, that alignment ceases. Consequently, the ${\bf J}\times{\bf B}$ force 
    becomes non-zero, acting on the MFR and rotating it (see Figure 11 of \citealt{2007ApJ...670.1453I}). As seen in Fig. \ref{f5}(e), the MFR's 
    total current increases and, TC1 in particular, roughly maintains its value up to $t \approx 5.3$. Therefore, it might contribute to the
    writhing motion. 
    Furthermore, \cite{2012SoPh..281..137K}  discussed  the writhing motion due to the relaxation of the twist during the eruption. They
    reported that, due to the relaxation, the magnetic axis of the MFR deviated from its initial position by about 40 degrees(see Figure 13 of 
    their paper). This effect would also contribute to the writhing motion of the MFR found in this study.  Consequently, the combination of 
    some effects might contribute to the writhing motion.  We wish to address detailed analysis of this as future work. 
    
    To discuss the initiation of the eruption, in Fig. \ref{f7}, we show the magnetic fields along with upward velocity ($V_z$) immediately after
    the initiation of the simulation. Note that the MFR located in the center corresponds to the FR1 discussed in Fig.\ref{f3}. The red, black, 
    and blue lines correspond to contours of the decay index ($n$) with values of 1.7, 1.5 and 1.0, respectively. The decay index $n$ is 
    defined as $n= -(z/B_{\rm ex})(\partial B_{\rm ex}/\partial z)$(\citealt{2006PhRvL..96y5002K})  where $B_{\rm ex}$ denotes the horizontal 
    component of external fields. Here we assume the potential field as the external field. We found that the upward velocity rapidly grows at 
    the region where n ranges from 1.0 to 1.7. Although the value of $n=1.5$ is well known as the threshold of the torus 
    instability(TI, \citealt{2006PhRvL..96y5002K}), the value strongly depends on the configuration of the MRF and line-tying effect({\it e.g.}, 
    \citealt{2010ApJ...718.1388D}, \citealt{2010ApJ...718..433O}, \citealt{2015Natur.528..526M}), and depending on the configuration the 
    eruptions can occur even when n is less or greater than 1.5.(\citealt{2015ApJ...814..126Z}, \citealt{2017ApJ...850...95S}). Therefore, TI 
    would be one of possible candidates to explain the initiation. On the other hand, as we discussed in previous section, since the 
    reconnection takes place in an early time of the simulation above which the FR1 starts to rise upward while expanding, it might play an 
    important role in the initiation. For instance, this reconnection might "push" the MFR in the region where it becomes torus unstable. In 
    order to draw a conclusion on the initiation of the eruption, therefore, a more detailed analysis is required. In addition, it is important to 
    understand the formation process of the magnetic fields producing the strong solar flares. In order to do, we need the temporal evolution 
    of the NLFFF  as shown in \cite{2016PASJ...68..101K}, \cite{2018arXiv180701436M} as well as detailed data 
    analysis(\citealt{2017ApJ...838..134B},  \citealt{2018ApJ...860..163W}).
                          
    The MFR shown in the simulation showed eastward deflection in the early stages as well as writhing motion.This might be important  
    to explain the magnetic fields observed in vicinity of Earth.  Because the AR was located 35 degrees west in longitude, this deflection 
    toward Earth could result in a strong field part of the MFR passing the Earth's position. Furthermore, the writhing motion found in the lower 
    corona might contribute to create the southward magnetic field observed in the vicinity of Earth. Therefore, the writhing motion might be 
    important in terms of space weather forecasting.
    
    Since this simulation is limited, as future work, extended simulations are required covering a larger area to take into consideration the 
    effects of ambient  field(\citealt{2010ApJ...718.1305S}). We also plan to examine a connection to the SUSANOO simulation, which is a 
    global solar wind model including CME propagation, developed by  \cite{2016SpWea..14...56S} toward a comprehensive understanding 
    of the evolution of the MFR from Sun to Earth.

\acknowledgments
 We are grateful to anonymous referee for helping us improve and  polish this paper. We would like to thank Prof. Kanya 
 Kusano, Dr.\ Bernhard Kliem, and Dr. \ Antonia Savcheva, for useful discussions. We are grateful  to Mr.\ Magnus Woods for checking this 
 manuscript. This study was motivated by useful discussions in 2nd PSTEP(http://www.pstep.jp/?lang=en) modeling workshop. {\it SDO} is a 
 mission of NASA's Living With a Star Program.This work was supported by JSPS KAKENHI Grant Numbers JP15H05814 and MEXT as 
 "Exploratory Challenge on Post-K Computer" (Environmental Variations of Planets in the Solar System).  This work was supported by the 
 computational  joint research program of the Institute for Space-Earth Environmental Research (ISEE), Nagoya University. The visualization
  was done by VAPOR(\citealt{2005SPIE.5669..284C}, 
 \citealt{2007NJPh....9..301C}).


\begin{thebibliography}{}
\bibitem[Amari et al.(1999)]{1999ApJ...518L..57A} Amari, T., Luciani, J.~F., Mikic, Z., \& Linker, J.\ 1999, \apjl, 518, L57 
\bibitem[Amari et al.(2003)]{2003ApJ...585.1073A} Amari, T., Luciani, J.~F., Aly, J.~J., Mikic, Z., \& Linker, J.\ 2003, \apj, 585, 1073
\bibitem[Amari et al.(2018)]{2018Natur.554..211A} Amari, T., Canou, A., Aly, J.-J., Delyon, F., \& Alauzet, F.\ 2018, \nat, 554, 211 
\bibitem[Aulanier(2014)]{2014IAUS..300..184A} Aulanier, G.\ 2014, Nature of Prominences and their Role in Space Weather, 300, 18 
{2010ApJ...718.1388D}
\bibitem[Bamba et al.(2017)]{2017ApJ...838..134B} Bamba, Y., Inoue, S., Kusano, K., \& Shiota, D.\ 2017, \apj, 838, 134
\bibitem[Benz(2017)]{2017LRSP...14....2B} Benz, A.~O.\ 2017, Living Reviews in Solar Physics, 14, 2
\bibitem[Berger \& Prior(2006)]{2006JPhA...39.8321B} Berger, M.~A., \& Prior, C.\ 2006, Journal of Physics A Mathematical General, 39, 8321 
\bibitem[Bobra et al.(2014)]{2014SoPh..289.3549B} Bobra, M.~G., Sun, X., Hoeksema, J.~T., et al.\ 2014, \solphys, 289, 3549 
\bibitem[Carmichael(1964)]{1964NASSP..50..451C} Carmichael, H.\ 1964, NASA Special Publication, 50, 451 
\bibitem[Chen(2017)]{2017PhPl...24i0501C} Chen, J.\ 2017, Physics of Plasmas, 24, 090501
\bibitem[Clyne \& Rast(2005)]{2005SPIE.5669..284C} Clyne, J., \& Rast, M.\ 2005, \procspie, 5669, 284 
\bibitem[Clyne et al.(2007)]{2007NJPh....9..301C} Clyne, J., Mininni, P., Norton, A., \& Rast, M.\ 2007, New Journal of Physics, 9, 301 
\bibitem[Dedner et al.(2002)]{2002JCoPh.175..645D} Dedner, A., Kemm, F., Kr{\"o}ner, D., et al.\ 2002, Journal of Computational Physics, 175, 645
\bibitem[D{\'e}moulin \& Aulanier(2010)]{2010ApJ...718.1388D} D{\'e}moulin, P., \& Aulanier, G.\ 2010, \apj, 718, 1388
\bibitem[Forbes(2000)]{2000JGR...10523153F} Forbes, T.~G.\ 2000, \jgr, 105, 23153
\bibitem[Gibson \& Fan(2006)]{2006ApJ...637L..65G} Gibson, S.~E., \& Fan, Y.\ 2006, \apjl, 637, L65 
\bibitem[Guo et al.(2017)]{2017ScChE..60.1408G} Guo, Y., Cheng, X., \& Ding, M.\ 2017, Science in China Earth Sciences, 60
\bibitem[Hirayama(1974)]{1974SoPh...34..323H} Hirayama, T.\ 1974, \solphys, 34, 323
\bibitem[Hood \& Priest(1979)]{1979SoPh...64..303H} Hood, A.~W., \& Priest, E.~R.\ 1979, \solphys, 64, 303 
\bibitem[Inoue et al.(2014a)]{2014ApJ...780..101I} Inoue, S., Magara, T., Pandey, V.~S., et al.\ 2014, \apj, 780, 101
\bibitem[Inoue et al.(2014b)]{2014ApJ...788..182I} Inoue, S., Hayashi, K., Magara, T., Choe, G.~S., \& Park, Y.~D.\ 2014, \apj, 788, 182
\bibitem[Inoue et al.(2015)]{2015ApJ...803...73I} Inoue, S., Hayashi, K., Magara, T., Choe, G.~S., \& Park, Y.~D.\ 2015, \apj, 803, 73
\bibitem[Inoue(2016)]{2016PEPS....3...19I} Inoue, S.\ 2016, Progress in Earth and Planetary Science, 3, 19
\bibitem[Inoue et al.(2018)]{2018NatCo...9..174I} Inoue, S., Kusano, K., B{\"u}chner, J., \& Sk{\'a}la, J.\ 2018, Nature Communications, 9, 174
\bibitem[Isenberg \& Forbes(2007)]{2007ApJ...670.1453I} Isenberg, P.~A., \& Forbes, T.~G.\ 2007, \apj, 670, 1453 
\bibitem[Janvier et al.(2015)]{2015SoPh..290.3425J} Janvier, M., Aulanier, G., \& D{\'e}moulin, P.\ 2015, \solphys, 290, 3425 
\bibitem[Jiang et al.(2016)]{2016NatCo...711522J} Jiang, C., Wu, S.~T., Feng, X., \& Hu, Q.\ 2016, Nature Communications, 7, 11522
\bibitem[Kang et al.(2016)]{2016PASJ...68..101K} Kang, J., Magara, T., Inoue, S., Kubo, Y., \& Nishizuka, N.\ 2016, \pasj, 68, 101
\bibitem[Kay et al.(2017)]{2017SoPh..292...78K} Kay, C., Gopalswamy, N., Xie, H., \& Yashiro, S.\ 2017, \solphys, 292, 78 
\bibitem[Kilpua et al.(2017)]{2017LRSP...14....5K} Kilpua, E., Koskinen, H.~E.~J., \& Pulkkinen, T.~I.\ 2017, Living Reviews in Solar Physics, 14, 5
\bibitem[Kliem \& T{\"o}r{\"o}k(2006)]{2006PhRvL..96y5002K} Kliem, B., \& T{\"o}r{\"o}k, T.\ 2006, Physical Review Letters, 96, 255002
\bibitem[Kliem et al.(2012)]{2012SoPh..281..137K} Kliem, B., T{\"o}r{\"o}k, T., \& Thompson, W.~T.\ 2012, \solphys, 281, 137 
\bibitem[Kopp \& Pneuman(1976)]{1976SoPh...50...85K} Kopp, R.~A., \& Pneuman, G.~W.\ 1976, \solphys, 50, 85 
\bibitem[Lemen et al.(2012)]{2012SoPh..275...17L} Lemen, J.~R., Title, A.~M., Akin, D.~J., et al.\ 2012, \solphys, 275, 17
\bibitem[Moore et al.(2001)]{2001ApJ...552..833M} Moore, R.~L., Sterling, A.~C., Hudson, H.~S., \& Lemen, J.~R.\ 2001, \apj, 552, 833
\bibitem[Muhamad et al.(2017)]{2017ApJ...842...86M} Muhamad, J., Kusano, K., Inoue, S., \& Shiota, D.\ 2017, \apj, 842, 86
\bibitem[Muhamad et al.(2018)]{2018arXiv180701436M} Muhamad, J., Kusano, K., Inoue, S., \& Bamba, Y.\ 2018, arXiv:1807.01436  
\bibitem[Myers et al.(2015)]{2015Natur.528..526M} Myers, C.~E., Yamada, M., Ji, H., et al.\ 2015, \nat, 528, 526
\bibitem[Olmedo \& Zhang(2010)]{2010ApJ...718..433O} Olmedo, O., \& Zhang, J.\ 2010, \apj, 718, 433 
\bibitem[Park et al.(2013)]{2013ApJ...778...13P} Park, S.-H., Kusano, K., Cho, K.-S., et al.\ 2013, \apj, 778, 13 
\bibitem[Park et al.(2018)]{2018SoPh..293....114} Park, S.~H et al.\ 2018, {\it Sol.\ Phys.}, 293, 114 
\bibitem[Pesnell et al.(2012)]{2012SoPh..275....3P} Pesnell, W.~D., Thompson, B.~J., \& Chamberlin, P.~C.\ 2012, {\it Sol.\ Phys.}, 275, 3  
\bibitem[Prasad et al.(2018)]{2018ApJ...860...96P} Prasad, A., Bhattacharyya, R., Hu, Q., Kumar, S., \& Nayak, S.~S.\ 2018, \apj, 860, 96 
\bibitem[Priest \& Forbes(2002)]{2002A&ARv..10..313P} Priest, E.~R., \& Forbes, T.~G.\ 2002, \aapr, 10, 313 
\bibitem[Romano et al.(2003)]{2003SoPh..214..313R} Romano, P., Contarino, L., \& Zuccarello, F.\ 2003, \solphys, 214, 313 \bibitem[Sakurai(1982)]{1982SoPh...76..301S} Sakurai, T.\ 1982, \solphys, 76, 301
\bibitem[Scherrer et al.(2012)]{2012SoPh..275..207S} Scherrer, P.~H., Schou, J., Bush, R.~I., et al.\ 2012, {\it Sol.\ Phys.}, 275, 207
\bibitem[Schmieder et al.(2015)]{2015SoPh..290.3457S} Schmieder, B., Aulanier, G., \& Vr{\v s}nak, B.\ 2015, \solphys, 290, 3457 \bibitem[Shiota et al.(2010)]{2010ApJ...718.1305S} Shiota, D., Kusano, K., Miyoshi, T., \& Shibata, K.\ 2010, \apj, 718, 1305
\bibitem[Shiota \& Kataoka(2016)]{2016SpWea..14...56S} Shiota, D., \& Kataoka, R.\ 2016, Space Weather, 14, 56 
\bibitem[Sturrock(1966)]{1966Natur.211..695S} Sturrock, P.~A.\ 1966, \nat, 211, 695
\bibitem[Syntelis et al.(2017)]{2017ApJ...850...95S} Syntelis, P., Archontis, V., \& Tsinganos, K.\ 2017, \apj, 850, 95
\bibitem[T{\"o}r{\"o}k et al.(2004)]{2004A&A...413L..27T} T{\"o}r{\"o}k, T., Kliem, B., \& Titov, V.~S.\ 2004, \aap, 413, L27
\bibitem[T{\"o}r{\"o}k et al.(2010)]{2010A&A...516A..49T} T{\"o}r{\"o}k, T., Berger, M.~A., \& Kliem, B.\ 2010, \aap, 516, A49 
\bibitem[van Ballegooijen \& Martens(1989)]{1989ApJ...343..971V} van Ballegooijen, A.~A., \& Martens, P.~C.~H.\ 1989, \apj, 343, 971
\bibitem[Wang et al.(2017)]{2017NatCo...8.1330W} Wang, W., Liu, R., Wang, Y., et al.\ 2017, Nature Communications, 8, 1330 
\bibitem[Wiegelmann et al.(2006)]{2006SoPh..233..215W} Wiegelmann, T., Inhester, B., \& Sakurai, T.\ 2006, \solphys, 233, 215
\bibitem[Wiegelmann \& Sakurai(2012)]{2012LRSP....9....5W} Wiegelmann, T., \& Sakurai, T.\ 2012, Living Reviews in Solar Physics, 9, 5
\bibitem[Woods et al.(2018)]{2018ApJ...860..163W} Woods, M.~M., Inoue, S., Harra, L.~K., et al.\ 2018, \apj, 860, 163 
\bibitem[Williams et al.(2005)]{2005ApJ...628L.163W} Williams, D.~R., T{\"o}r{\"o}k, T., D{\'e}moulin, P., van Driel-Gesztelyi, L., \& Kliem, B.\ 2005, \apjl, 628, L163
\bibitem[Yan et al.(2014)]{2014ApJ...782...67Y} Yan, X.~L., Xue, Z.~K., Liu, J.~H., et al.\ 2014, \apj, 782, 67
\bibitem[Yan et al.(2018)]{2018ApJ...856...79Y} Yan, X.~L., Wang, J.~C., Pan, G.~M., et al.\ 2018, \apj, 856, 79
\bibitem[Yang et al.(2017)]{2017ApJ...849L..21Y} Yang, S., Zhang, J., Zhu, X., \& Song, Q.\ 2017, \apjl, 849, L21
\bibitem[Zuccarello et al.(2015)]{2015ApJ...814..126Z} Zuccarello, F.~P., Aulanier, G., \& Gilchrist, S.~A.\ 2015, \apj, 814, 126\end{thebibliography}
\end{document}